\documentclass[runningheads]{llncs}

\usepackage[english]{babel}
\usepackage{blindtext}
\usepackage{csquotes}
\usepackage[useregional]{datetime2}
\usepackage[obeyspaces]{url}
\usepackage[caption=false]{subfig}
\usepackage{refcount}
\usepackage[sort]{cite}

\usepackage{amssymb}
\usepackage{tabularx}
\usepackage{booktabs}

\def\clap#1{\hbox to 0pt{\hss#1\hss}}

\usepackage{xspace}
\newcommand*{\eg}{e.g.\@\xspace}

\usepackage{tikz}
\usetikzlibrary{positioning}
\usetikzlibrary{scopes}
\usetikzlibrary{calc}
\usetikzlibrary{decorations.pathmorphing}
\usetikzlibrary{decorations.pathreplacing}
\usetikzlibrary[shapes.symbols]

\begin{document}
\title{Operating Tor Relays at Universities: Experiences and Considerations}
\titlerunning{Operating Tor Relays at Universities}

\author{Christoph Döpmann\inst{2} \and
Matthias Marx\inst{1} \and
Hannes Federrath\inst{1} \and
Florian Tschorsch\inst{2}}

\authorrunning{Döpmann et al.}
\institute{University of Hamburg, Germany\\
\email{\{marx,federrath\}@informatik.uni-hamburg.de}
\and
Technische Universität Berlin, Germany\\
\email{\{christoph.doepmann,florian.tschorsch\}@tu-berlin.de}}

\maketitle

\begin{abstract}
In today's digital society,
the Tor network has become an indispensable tool for individuals
to protect their privacy on the Internet.
Operated by volunteers, relay servers constitute the core component of Tor
and are used to geographically escape surveillance.
It is therefore essential to have a large, yet diverse set of relays.
In this work, we analyze the contribution of educational institutions to the Tor network
and report on our experience of operating exit relays at a university.
Taking Germany as an example (but arguing that the global situation is similar),
we carry out a quantitative study
and find that universities contribute negligible amounts of relays and bandwidth.
Since many universities all over the world have excellent conditions
that render them perfect places to host Tor (exit) relays,
we encourage other interested people and institutions to join.
To this end, we discuss and resolve common concerns
and provide lessons learned.

\end{abstract}

\section{Introduction} \label{sec:introduction}

With about 6,800 relays and 2~million daily users~\cite{tor-metrics},
the Tor network~\cite{dingledine2004tor} is today's most popular anonymization network.
Anonymity is realized by relaying data over a cryptographically secure path of relays,
where each hop knows its neighbors only.
Relays are contributed and administrated by volunteers,
often running on their private or leased systems.

In this paper%
\footnote{This article is based on our German paper published in~\cite{dud-paper}.},
we report on our experiences of running a Tor exit relays at a universities.
During our research in the area of anonymity networks in general and Tor in particular,
we were faced with concerns about running a Tor relay.
In the following, we evaluate the pros and cons to foster a discussion on the topic.
At the same time, we intend to motivate interested people
to make their own experiences of operating a Tor relay.

In fact, universities denote prime candidates for hosting Tor relays
as they have good network connectivity,
assemble the technical expertise to run relays,
and generally value freedom of thought and expression.
Hence, we believe that universities could strengthen the Tor network.
Taking Germany as an example, however,
only two Tor exit relays are operated by universities as of today.

We discuss why it would be desirable if more universities joined,
improving relay diversity.
To this end, we discuss and aim to resolve \enquote{typical} concerns
of setting up and operating Tor relays,
putting a special focus on exit relays.
By explaining how the Tor network works and how we solved issues,
we do not only address fellow researchers and regular users with similar problems,
but also providers or co-location center operators at universities.
Please note, however, we are \emph{not} focusing on legal aspects;
our intention is rather to report independent from laws,
which should help to understand the design concepts of the Tor network regarding its risks.
Our work shows that while challenges exist,
they are fully manageable.

This paper is organized as follows.
Section~\ref{sec:rel-work} presents related work and provides an explanation of how Tor relays work.
In Section~\ref{sec:vorstudie}, we quantify the amount of resources
that universities have contributed to the Tor network, both globally and especially in Germany.
In doing so, we provide qualified arguments as to why universities and the research community
are suitable places for operating Tor (exit) relays.
In Section~\ref{sec:case-studies}, we then report on our own experience
of operating two Tor exit relays at German universities, both technically and organizationally.
We elaborate on the situation at our universities in Section~\ref{sec:case-studies},
sharing our experiences.
We leverage this insight in Section~\ref{sec:conclusion} to discuss typical concerns
and develop a checklist helping universities to operate Tor relays,
before Section~\ref{sec:conclusion} concludes the paper.

\section{Background and Related Work} \label{sec:rel-work}

\subsection{Tor Relays}

Relays denote the core component of the Tor network~\cite{dingledine2004tor}.
As such, their main task consists in forwarding traffic through the network,
where traffic is divided into chunks of data, called \emph{cells}.
Typically three relays are combined into a \emph{circuit} by the clients.
Since there can be many concurrent clients, relays handle many circuits at a time.
By construction, a relay knows its immediate predecessor and successor only,
which is necessary to ensure anonymity.
To this end, Tor makes use of the principle of \emph{onion routing}~\cite{goldschlag1996onionrouting},
which involves cryptographic operations when forwarding cells.
In particular, for each circuit $C$, a relay $R$
shares a symmetric cryptographic key $K_{R,C}$ with the circuit’s client,
securely exchanged during circuit construction.
This key is used during the forwarding of traffic
for carrying out the main cryptographic operation, adding and removing layers of encryption.
A cell from the client has been encrypted multiple times,
with all $K_{x,C}$ for each relay $x$ in the circuit.
When forwarding the cell towards the other end of the circuit (the \emph{exit}),
every relay removes its layer of encryption by decrypting the cell with its symmetric key.
Vice versa, for cells traveling the other direction,
a relay encrypts the cell using its key, effectively adding a layer of encryption.
Therefore, contents of a cell cannot be read or tracked over relays.

\begin{figure}[htp]
\centering
\subfloat[The first relay is the only relay that accepts direct connections from clients and is called the \emph{entry guard}. It routes data between client and middle relay.\label{fig:roles:guard}]{%
  \includegraphics[width=0.9\textwidth]{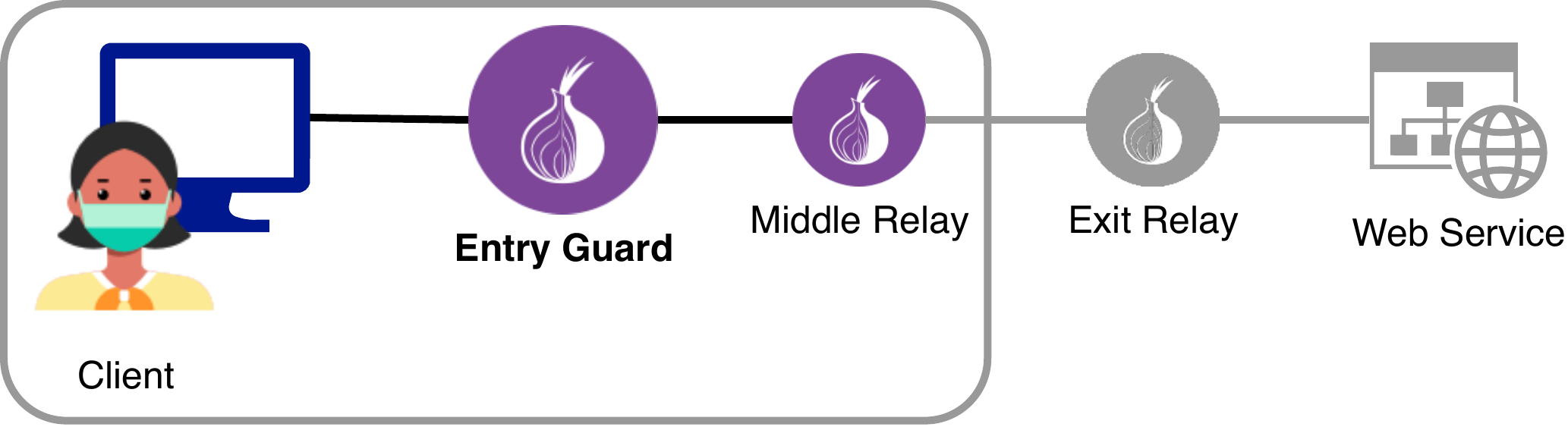}%
}\\
\subfloat[The \emph{exit} relay constitutes the \enquote{end} of a circuit. It routes data between middle relay and target web service.
\label{fig:roles:exit}]{%
  \includegraphics[width=0.9\textwidth]{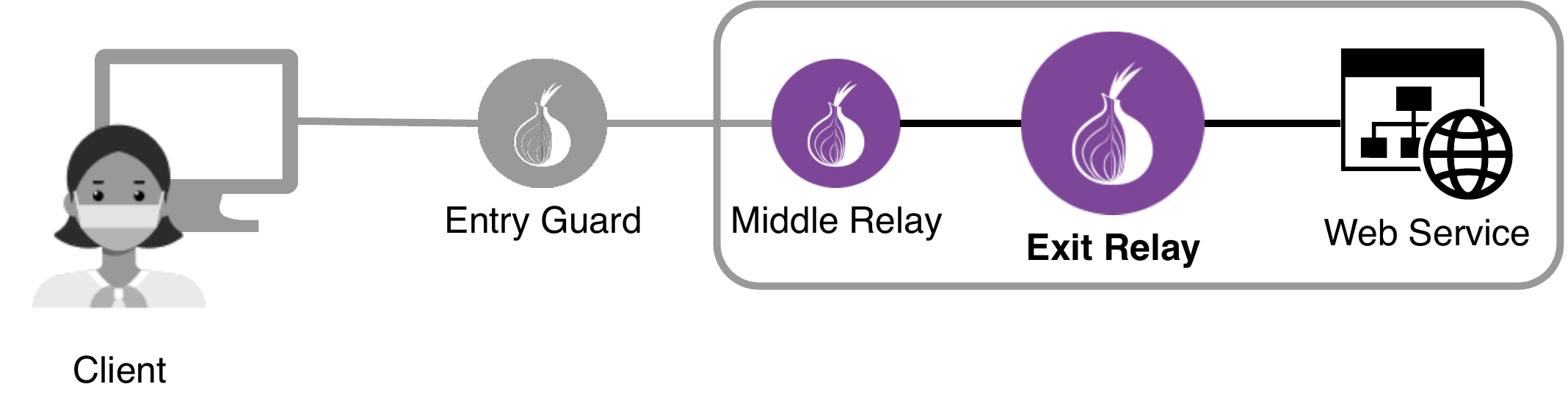}%
}

\caption{Three relays are combined into a \emph{circuit} by a client.
By construction, a relay knows its immediate predecessor and successor only. This is illustrated here for \protect\subref{fig:roles:guard}~\emph{Entry Guards} and \protect\subref{fig:roles:exit}~\emph{Exit Relays}.}
\label{fig:roles}
\end{figure}

Depending on the position in a circuit, relays can take different roles,
which are depicted in Figure~\ref{fig:roles}.
Two roles stand out:
The first relay is the only relay that accepts direct connections from clients and is  called the \emph{entry guard} (see Figure~\ref{fig:roles:guard}).
While every relay can take the entry role,
it needs to satisfy several conditions such as a high degree of availability and bandwidth
to receive the \enquote{guard} flag.
An entry guard operator can observe the IP addresses of Tor clients but does not know the actual target servers.
On the other hand, the \emph{exit} relay constitutes the \enquote{end} of a circuit
and is therefore especially important
as it enables anonymous access to the Internet.
In particular, the client instructs the exit relay
to establish TCP connections to target servers on the Internet
as well as sending and receiving data from these connections.
An exit node operator can observe connections to the target servers
but does not know the IP addresses of the corresponding Tor clients.
Exit relays are accordingly the nodes that are visible to external servers
and may appear as clients in their logs (see Figure~\ref{fig:roles:exit}).
If the Tor network is misused, for example to attack a web server,
then it looks like the exit relay is attacking the server.
As this behavior may not be desired for relay operators, relays do not act as exits by default.
In contrast, the exit role must be activated deliberately by the relay operator.
In the past, this has resulted in exit capacity being a scarce resource~\cite{jansen2019pointbreak}.
Operators may also restrict the set of IP addresses and remote ports
circuits are allowed to connect to via this exit relay.
This set of rules is known as \emph{exit policy}.

For another reason, the exit relay plays a special role.
Due to the onion encryption scheme, it sees the client data in plaintext.
In order to protect confidentiality and integrity,
clients need to employ end-to-end encryption and authentication schemes
such as TLS.

In addition to anonymous Internet access,
\emph{Onion Services} offer Tor-internal (and therefore anonymous) services.
To this end, a client's circuit is connected to the service's circuit
via a so-called rendezvous point.
Accordingly, the traffic never leaves the Tor network.

\subsection{Related Work}

There are a number of guides and reports that provide information on operating Tor (exit) relays in general
and relay operation at universities in particular.
The Tor Project itself provides information on the operation of exit relays~\cite{tor-tips-exit-relays}.
The community has also compiled recommendations specifically for universities~\cite{tor-community-relay-universities,tor-community-universities-isp}.
From these documents, we can learn that it is recommend to familiarize with the university's acceptable use policy, local laws with respect to liability of traffic, design of Tor, and authentication on the Internet.
Future operators should also find allies and teach the university's lawyers and network security people about Tor. If abuse messages are a problem, it is recommended to use a descriptive host name, adjust the relay's bandwidth, or to reduce the exit policy. Similar information is provided by the Electronic Frontier Foundation~(EFF), but in a broader context~\cite{eff-tor-on-campus}.

In 2011, Thomas Lowenthal shared his experiences on running Tor exits at Princeton University~\cite{lowenthal-princeton}. In 2014, Jesse Victors shared his experiences from Utah State University~\cite{victors-utah-state} and the EFF published a collection of stories by various universities~\cite{eff-tor-university-experiences}.
Beyond Tor, there are also reports on the operation of other anonymization services,
which involve similar concerns.
The project partners who developed the Java Anon Proxy, later known as JonDonym, shared their experiences in operating an anonymizing service for the period from~2000 to 2010~\cite{golembiewski2003experiences,kopsell2003erfahrungen,kopsell2010entwicklung}. They also shared their experience with law enforcement agencies and a court order in 2003~\cite{baeumler2003}.

With more than six years passed by since the last publication, we think it is time again to report on current experiences of running Tor relays in an university environment.

\section{Tor Relays at Universities} \label{sec:vorstudie}

The Tor network consists of around 1,200 exit relays of a total of 6,800 relays that are run by volunteers.
There is a whole ecosystem around Tor, including its users and a research community.
Although the Tor network is researched extensively, universities provide a fraction of relays and bandwidth only.
In this Section, we quantify this circumstance.
We first focus on Germany as an example due to its significance for the Tor network.
Afterwards, we give preliminary arguments why this denotes a representative example for the whole network.

\subsection{Situation in Germany} \label{sec:vorstudie-deutschland}

We mainly focus on Germany as an example for several reasons:
First, Germany is the single most important home of relays,
comprising more than 30\% of the Tor network's capacity,
which is more than double the second country (France)~\cite{tor-metrics}.
Second, Germany is generally regarded to have a relatively privacy-conscious society,%
also giving strong civil rights to educational institutions.
One would therefore expect a large number of exit relays being also operated at German universities.
And finally, the Internet access of German universities is more homogeneous
than in other countries,
because the vast majority of universities uses the same public non-profit ISP
called \emph{Deutsches Forschungsnetz} (DFN, AS~680),
which has its own IP address space and autonomous system.
Therefore, Tor relays can easily and reliably be classified
as either educational or non-educational.

In order to assess the situation of Tor relays at universities in Germany,
we first focus on exit relays and take the following metrics into account:
\begin{itemize}
\item overall exit bandwidth in the Tor network
\item exit bandwidth of relays in Germany
\item exit bandwidth of relays in DFN (AS~680)
\end{itemize}
We do not only consider today's situation,
but also the historic development.
We therefore processed historic Tor relay server descriptors
and rely on the MaxMind database~\cite{maxmind} for IP geolocation and AS information.

As of today, German relays provide about 30\% of Tor's exit bandwidth.
However, German universities account only for 0.2\% of the global exit bandwidth.
13 relays, including two exit relays, are operated within the DFN.
In contrast, there are associations and individuals
who contribute between 1\% and 20\% of the exit bandwidth~\cite{nusenu-exit-families}.

\begin{figure}[t]
\centering
\subfloat[Number of relays.\label{fig:relay-stats:relays}]{%
  \includegraphics[width=0.49\textwidth]{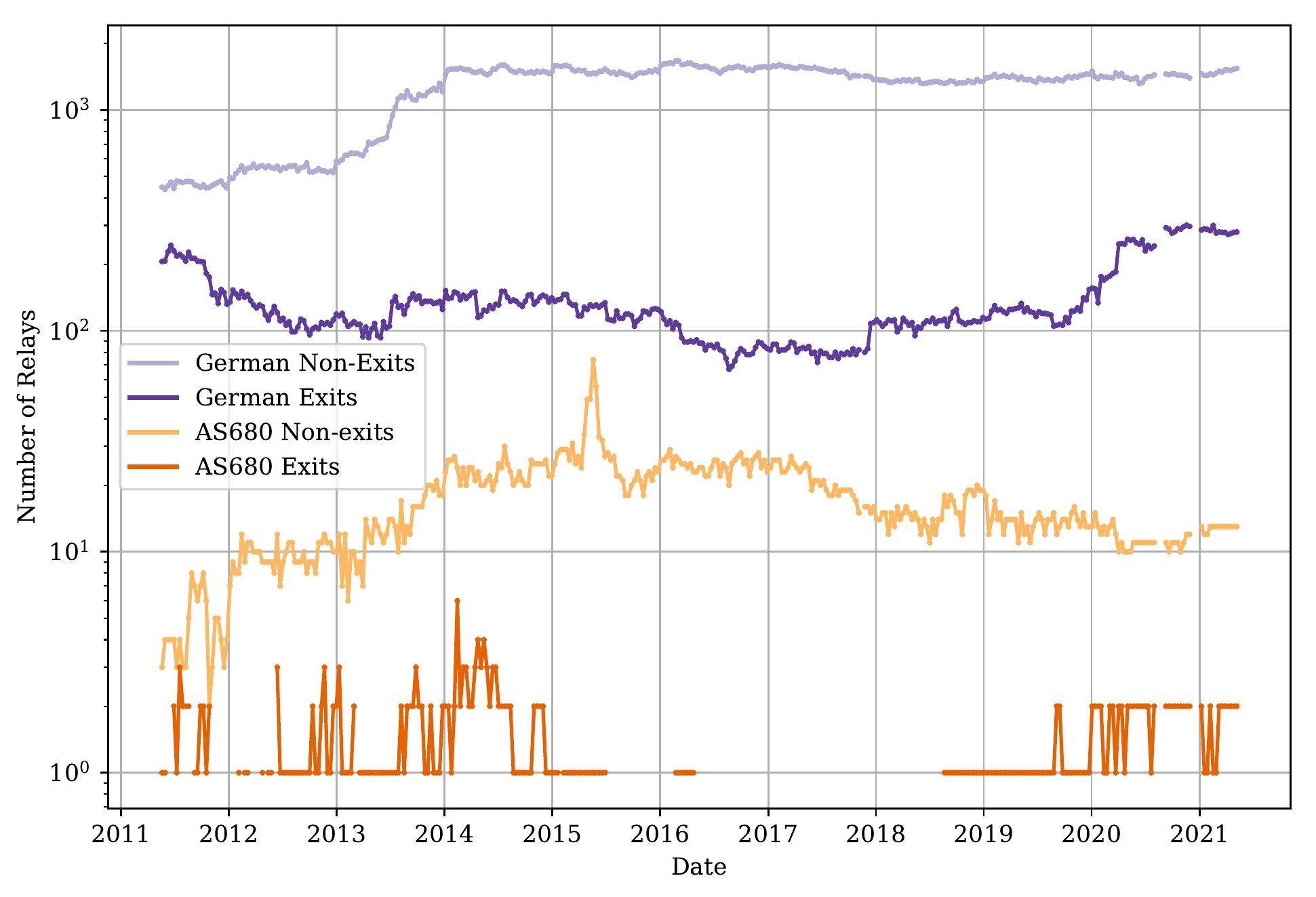}%
}\hfill
\subfloat[Relay bandwidth.\label{fig:relay-stats:bandwidth}]{%
  \includegraphics[width=0.49\textwidth]{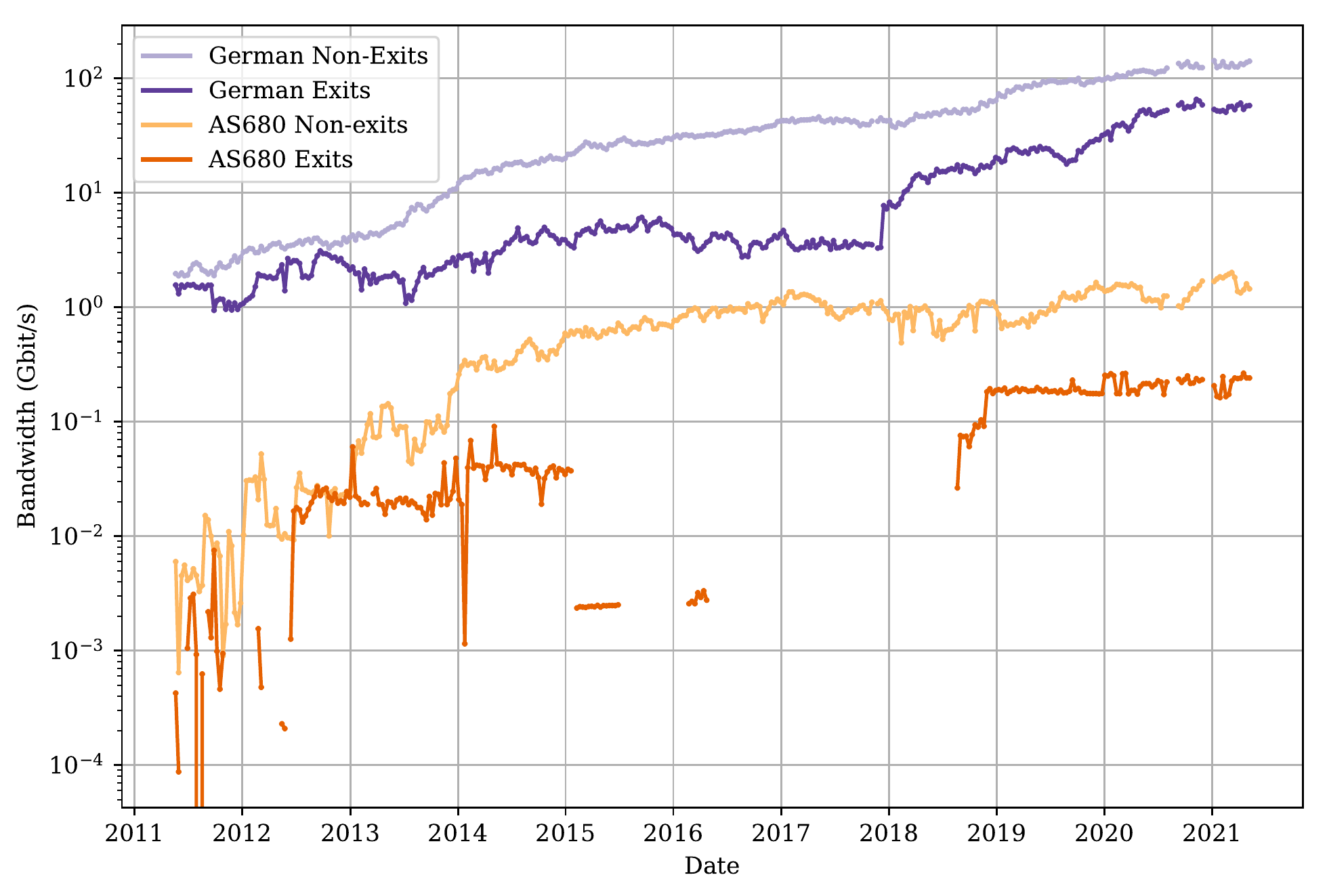}%
}

\caption{Time series for the past ten years for \protect\subref{fig:relay-stats:relays}~the number of relays and \protect\subref{fig:relay-stats:bandwidth}~the combined relay bandwidth in ASN 680 (Deutsches Forschungsnetz) and in Germany. %
}
\label{fig:relay-stats}
\end{figure}

Taking a look at the historic trend,
one can see that this has been similar in the past.
Figure~\ref{fig:relay-stats} shows the bandwidth as well as the number of relays
for exits and non-exits in Germany over the last 10~years.
Since 2014, the number of non-exit relays in Germany
has been relatively constant at about 1,400.
The number of exit relays in Germany has experienced more fluctuation,
peaking in 2011~(244) and 2020~(261),
but dropping to 71~relays in 2016.
The provided bandwidth exhibits a constant upward trend, though.

The proportion of relays at German universities has always been small,
both in terms of bandwidth and number of relays.
At most 73~non-exit relays have been operated at universities (2015).
Since then, the number has fallen to 13.
For exit relays, the numbers are even clearer.
In the last ten years, there has been a maximum of five exit relays at universities in Germany,
mostly one or two, or even none at all for long periods of time.
Consequently, the bandwidth has long been negligible.
Since August 2018, there has been at least one exit relay (operated by TU Berlin),
providing higher bandwidth than before.
Since January 2019, with interruptions, there has been a second relay (operated by University of Hamburg).
However, we can conclude that universities in Germany
contribute only little to the Tor network.

We believe that universities are a suitable place to operate Tor relays.
This is not only because Tor is subject of research and education.
Universities can also contribute as places of academic freedom
(freedom of research, teaching and study)
to make it possible for others to circumvent censorship and to express freedom of speech.
Finally, universities are well-suited for technical reasons.
They often have their own IP address space,
which is useful for avoiding conflicts in case of abuse complaints.
Moreover, they usually have a good and reliable connection to the Internet
that can handle some additional bandwidth for Tor.
Therefore, it would be desirable to see more Tor (exit) relays
being hosted at universities in the future.

\subsection{Global Situation}

While our main focus is on Germany,
we provide preliminary evidence that this situation may indeed be representative
also for the whole Tor network, on a global scale.
We cannot carry out the same, precise measurement as in Section~\ref{sec:vorstudie-deutschland}
because not every country has a designated research network for universities,
which makes automated classification more difficult.
For this experiment, we instead rely on a commercial IP address classification service%
\footnote{\enquote{Company API} offered by \url{https://ipinfo.io/}}
that is capable of providing a user type for given IP addresses
(\eg, \enquote{company}, \enquote{hosting}, or \enquote{education}).
While we cannot guarantee any specific accuracy,
the results appear plausible to us
and we manually verified they did not contain false positives.

We classified all relays from a current snapshot of the Tor network
and found that, worldwide,
educational institutions contribute about 0.7\% of the exit bandwidth
(10 out of 1,178 exit relays; 0.8\%).
This aligns reasonably well with the German situation presented above
and shows that universities should contribute more to the Tor network.

\section{Experiences Operating Exit Relays}
\label{sec:case-studies}

We here give a brief impression of how we managed to set up
Tor exit relays at our universities.
As of today, these are the only two exit relays run at German universities.

\subsection{Case Study 1: Technische Universität Berlin}

Since August~2018, the research group on Distributed Security Infrastructure at TU Berlin
is operates a Tor exit relay\footnote{Fingerprint of the relay:
\texttt{E91905CFEB230B1BEA6B0309816F9EE9C1A1A83A}}.
We decided to set it up not only to strengthen the network,
but also to support our Tor research,
for example to gain an in-depth understanding of the network traffic dynamics in Tor.
Moreover, we teach courses that include the topic of anonymous communication systems
and the relay allows us to report first-handed experiences.
We had operated non-exit Tor relays at other German universities before,
but were faced with completely different obstacles.
However, this was our first exit relay.
The technical setup at TU Berlin is as follows:
The university operates a data center on campus,
offering server housing as well as hosting of virtual machines.
We installed a virtual machine with 2~CPU cores, 2~GB of RAM
as well as an Internet access link of 1~Gbit/s.
While there are potential downsides of virtualization in our scenario
(\eg affecting other virtual machines),
this is what our small working group could contribute.
The data center is connected to the Internet via the DFN backbone.
Note that, in general, it is recommended to run a Tor relay
on a dedicated server to separate it from other infrastructure.
However, this was not an option for our small research group
due to financial limitations.
However, thanks to a hardware donation arranged by Artikel10~e.V.,
we will be able to run our own physical servers in the near future.

We took care to properly configure the Tor software
to expose valid and working contact information.
For example, on port~80,
we display an information page explaining Tor and our research,
as well as contact details.
Our intention was to inform any third party
that may observe abusive traffic from our IP address, openly and directly.
In addition, this allows the Tor community to see that our relay
is subject to research.
We use one of the recommended default exit policies
that balances usability with risk of abuse~\cite{tor-community-reduced-exit-policy}.
Furthermore, we decided to begin with a conservative bandwidth limit
to gain some experience first.
Later on, we increased the bandwidth limit to 160~Mbit/s,
which is still our current configuration.
We have observed that the hardware resources we have available (especially CPU and RAM)
do not allow to operate at higher data rates.
Forwarding larger amounts of traffic constantly
was generally not a concern at TU Berlin,
in contrast to other universities we have operated relays at.

Our experience from setup and operation is very positive.
We faced very few incidents that required our attention.
We were once alerted by DFN and TU Berlin's data center operator about
possible misuse of our server, which appeared to be serving malware.
We explained the situation and continued operation as normal.
When such an alert came up again, we seeked for a permanent solution
and asked to ignore similar warnings about our server in the future.
This was possible because DFN regards such warnings as a service
to its customers (TU Berlin in this case) and not as a strict demand for action.
Since then, we have not received any abuse complaint again.
Our contact email in the Tor directory does receive large amounts of spam,
but since we use a \mbox{separate} mailbox, this is not much of an issue.
The maintenance efforts are almost negligible.
In essence, they consist of upgrading to newly published software versions,
which could even be automated.

\subsection{Case Study 2: University of Hamburg}
Since January~2019, University of Hamburg operated a non-exit Tor relay. In September~2019 the Working Group on Security and Privacy began operating a Tor exit relay\footnote{Fingerprint of the relay:
\texttt{83C50784528AD3823CB7E7DF4B34B92A42CC7639}} after they became responsible for a small range of IP addresses. With its own IP range, the working group is also responsible for handling the corresponding abuse notifications. A mailing list has been set up so that several employees can receive and respond to the abuse notifications. A simple template for responses to abuse complaints has been prepared.

During preparation of the relay operation, many recommendations of the Tor project for the operation of (exit) relays in general (of which the working group was aware) and of relays at universities (of which the working group was not aware) were taken into account.
The university's acceptable use policy was known from other projects.

The Tor software runs on a dedicated server, the server is connected to the Internet via the DFN backbone. Ansible\footnote{\url{https://github.com/nusenu/ansible-relayor}} is used to manage the relay.

About once per day, the working group receives automated alerts from DFN's incident response team %
because malware appears to be running on the server while in fact malware is (ab)using the exit relay to connect to command and control servers.
Since these alerts do not require a response and the reason is known,
we do not take any further actions.

To keep the amount of abuse notifications low, the exit policy usually allows port~80 and 443 only. However, other ports were temporarily allowed during various experiments. Experiments are also the reason for various downtimes. It is noteworthy that the permission of port~22 (SSH) led to a significant increase in abuse notifications.

Most of the received abuse notifications were generated automatically (\eg, by Fail2Ban). Some did not have a valid reply address. A few asked to prevent future abuse, their IP addresses were then excluded from the exit policy. Most reports relate to brute-force attacks.

In one case in 2020, we were contacted by the police
that asked for our support in a process of investigation.
We informed the police that the IP address in question denotes a Tor exit relay,
and mapping the IP address to single users is technically impossible.
We also briefly explained current research projects involving the Tor relay.

\section{Considerations and Concerns}
\label{sec:considerations}

Setting up and running a Tor relay is an important way of strengthening the Tor network,
especially if done from within an educational institution.
Doing so should be thought-through thoroughly, though.
Based on our experience, %
we describe and assess some arguments and concerns that may be considered when
running Tor relays.

\subsection{Discussion of Concerns}

First, the network traffic a Tor relay generates can be substantial.
However, as a relay operator, one does have control over the amount of bandwidth
that is consumed.
The Tor software offers to configure upper bounds for the average and burst
bandwidth it is allowed to use.
While this is enforced locally, it is also published to Tor's directory service
used for relay discovery.
These so-called bandwidth weights are then taken into account by clients
to load balance the random relay selection during circuit construction.
After all, we believe that this is a way to safely maximize the donated bandwidth.

Another obstacle when setting up a Tor relay at universities
consists in ensuring proper reachability of the network service.
In particular, this includes configuring firewalls accordingly.
The administration staff of such corporate-like infrastructures
may have rather rigid firewall rules and little flexibility to adapt them.
In this regard, however, relay operators are offered quite some flexibility from the Tor software:
The TCP port used for relay operation (\texttt{ORPort}) can be chosen freely.
If possible, port 443 is recommended because it can be used by the most number of clients.
However, less regular ports like 9001 can just as well be used,
for example to avoid collision with other services.
The same is true for the directory port (\texttt{DirPort}) if the relay is configured as a directory cache.

In case of legal actions, it is often recommended that
a separate server as well as a separate IP address be used for the relay.
This is due to the fact that it greatly facilitates to
attribute any observed traffic to this service,
\eg, when being confronted with abuse complaints.
A separate IP address or subnet can also be beneficial,
as some email or other service providers blacklist IP addresses of Tor relays~\cite{singh2017characterizing}.

In general, it can be said that the configuration effort
for setting up a Tor relay is rather low,
also due to example resources offered by the Tor project.

When setting up a Tor relay, an important decision to take
is whether the relay should also act as an exit node.
From the technical point of view,
this requires the relay operator to allow arbitrary outgoing TCP connections from the relay.
When acting as an exit, though, public Internet hosts will see the relay’s IP address
as the client address for any requests relayed over this node.
As a consequence, any potential misuse may likely be wrongly attributed to the relay operator.
As practical experience has shown, anonymity also attracts illegal activities.
Relay operators therefore have to be prepared to handle abuse complaints
about Internet activities relayed by the Tor node.
However, the two cases of our exit relays have shown that, as of today,
the risk of being confronted with serious complaints appears to be rather low.
One important factor is the adoption of a suitable exit policy,
particularly one that disallows connections to SSH servers.
Properly configured, the number of complaints---%
and thus the efforts necessary to handle such requests---%
are manageable.
Comparing this to the situation in the past,
we suppose that a variety of factors have contributed
to the heavily reduced risk connected with operating an exit relay.
For example, the concept of Tor has become much more apparent to legal authorities.
Moreover, data protection regulation like the GDPR
has made it much harder to \emph{automatically} file complaints against relay operators,
\eg, due to reduced data in WHOIS responses.

\subsection{Lessons Learned}

\begin{table}
  \caption{Organizational checklist for operating Tor exit relays at universities.}
  \label{tab:checklist}
  \begin{tabularx}{\textwidth}{lX}
    \toprule
    $\square$ & \textbf{Talk to university staff.}\newline
    Get in touch
    with the people at your university that are concerned with hosting your server.
    In particular, this involves entities like the data center or network operator.
    Let them clearly know what you are doing
    and demonstrate that you know what you are doing.
    Explain to them the significance of hosting a Tor exit relay.
    Also, the freedom of research is an important value to base your undertaking upon
    if you are working in the field, so make use of it.\\
    \midrule
    $\square$ & \textbf{Organize and assign maintenance jobs.}\newline
    Running a Tor relay does not require much maintenance effort.
    If network operators are concerned about potential additional workload,
    this can be an encouragement.
    Offer to handle potential incoming abuse complaints.
    It is important to have responsibilities, processes, and templates ready for such cases:
    both, to have them smoothly handled, but also to reassure operators.
    You may also consider setting up unattended upgrades.\\
    \midrule
    $\square$ & \textbf{Communicate your intentions.}\newline
    Provide valid contact information that can be used to get in touch with you.
    For example, configure your exit relay to present an information website,
    publish your contact info to the Tor directory,
    set up whois data for your IP address (if possible),
    and configure a meaningful reverse DNS name.
    Running a Tor relay contributes to an open and free Internet.
    If you are confronted with rejection or lack of understanding, though,
    be prepared to explain why it is important.\\
    \midrule
    $\square$ & \textbf{Find allies at your university.}\newline
    Especially if you are a student, it will likely be very beneficial
    to reach out for research groups or university staff to support you.
    Raising awareness of the project can also be achieved
    by generally initiating or supporting discussions about
    the freedom of the Internet, privacy, and related topics.\\
    \midrule
    $\square$ & \textbf{Get in touch with the community.}\newline
    Ensure to stay up to date about changes and developments of the Tor network.
    The easiest way is to subscribe to the \emph{tor-announce} mailing list
    and potentially others.\footnote{%
    \emph{tor-relays} may be too noisy and \emph{tor-relays-universities}
    has been inactive for several years}
    You may also consider participating in Tor Relay Operators Meetups
    as a means of experience exchange.
    The meetups are often announced through the \emph{tor-relays} mailing list
    or the Tor Project website.\\
    \midrule
    $\square$ & \textbf{Teach about Tor.}\newline
    It is generally important to teach students as well as staff
    about the importance of Tor and how it works.
    However, in the context of operating your own relay,
    it becomes especially useful to create a technical understanding of Tor.
    Consider, for example, offering dedicated courses, projects, or seminars.
    You could even use the process of setting up the Tor relay as a project idea.\\
    \midrule
    $\square$ & \textbf{Point out benefits for the university.}\newline
    In order to get the university aboard (not only technical staff, but also
    management and heads), point out how operating a Tor relay aligns with
    the university's core values and principles such as freedom of speech
    and that operating a Tor relay is an excellent way to support these.\\
    \bottomrule
  \end{tabularx}
\end{table}

We leverage our experience from setting up Tor exit relays at universities
to compile a checklist that should support others following our footsteps
(see Table~\ref{tab:checklist}).
The list is based on several public sources~%
\cite{tor-community-relay-universities,eff-tor-on-campus,tor-new-relay-guide},
but we filter, aggregate, and augment their points
based on our experiences.
The list assumes that the operator acquired sufficient (technical)
knowledge of the Tor network
for setting up and configuring a relay.
The checklist is therefore meant as a list of \emph{action items}
that should be taken deliberately.

The checklist is compiled with the ultimate goal of operating an exit relay in mind,
because this is what the Tor network can benefit from the most.
However, it may be an option to start with a non-exit relay
and enable the exit functionality later,
when the operator feels confident enough to do so.
The checklist expresses our experience that responsibilities should be clarified
and it is important to get in touch with anyone who might be involved.

All in all, we come to the conclusion that typical technical and non-technical arguments
against running Tor (exit) relays at universities
can be addressed (or avoided) with reasonable effort.
In favor of running Tor relays speaks the possibility of using them for research and teaching.
Possible topics include anonymization techniques in general
or improvements of Tor in particular, Internet censorship,
as well as legal and ethical issues in the operation of anonymization networks.

\section{Conclusion} \label{sec:conclusion}

Our results show that, as of today, an increased involvement of universities
in supporting the Tor network with (exit) relays,
is not only reasonable and maybe even necessary,
but also absolutely possible.
Taking Germany as an example,
we demonstrate that the Tor network could be strengthened significantly
by stronger involvement of educational institutions.
Backed by our own experience of running Tor exit relays at universities,
we thus call upon universities to start operating more Tor (exit) relays.

\bibliographystyle{splncs04}
\bibliography{local.bib,bib/dsi.bib,bib/anon.bib,bib/blockchain.bib,bib/conferences-crossref.bib}

\footnotesize
\vspace{1em}
All websites were last accessed on \DTMDate{2021-05-03}.

\end{document}